\pdfoutput=1

\documentclass{svproc}
\usepackage{url}

\usepackage[ruled]{algorithm2e}
\usepackage{booktabs}
\usepackage{amsmath,amssymb}
\usepackage{enumerate}
\usepackage{graphicx}

\begin{document}
\mainmatter              
\title{Multilayer Block Models for Exploratory Analysis of Computer Event Logs}
\titlerunning{Multilayer Block Models for Exploratory Analysis of Computer
Event Logs} 
%
\author{Corentin Larroche}
\authorrunning{Corentin Larroche} 
%
\tocauthor{Corentin Larroche}
\institute{French National Cybersecurity Agency (ANSSI), Paris, France\\
\email{corentin.larroche@ssi.gouv.fr}
}

\maketitle              

\begin{abstract}
We investigate a graph-based approach to exploratory data analysis in the
context of network security monitoring.
Given a possibly large batch of event logs describing ongoing activity, we
first represent these events as a bipartite multiplex graph.
We then apply a model-based biclustering algorithm to extract relevant
clusters of entities and interactions between these clusters, thereby providing
a simplified situational picture.
We illustrate this methodology through two case studies addressing
network flow records and authentication logs, respectively.
In both cases, the inferred clusters reveal the functional roles of entities
as well as relevant behavioral patterns.
Displaying interactions between these clusters also helps uncover malicious
activity.
Our code is available at
\texttt{https://github.com/cl-anssi/MultilayerBlockModels}.
\end{abstract}
\section{Introduction}

The ever-increasing prevalence of cyber threats compels network
administrators and information security specialists to constantly
monitor ongoing activity inside sensitive networks.
Such monitoring often boils down to collecting and analyzing event logs, which
can be generally defined as structured records reflecting all sorts of actions.
These logs are continuously produced in massive amounts by nearly all devices
within a monitored network, which is both a blessing and a curse: while wide
and systematic log collection provides a detailed situational picture,
extracting actionable insights from such volumes of data is highly challenging.
Data mining algorithms are thus needed to efficiently analyze and summarize
all the available information.

Interestingly, various types of event logs can be naturally represented as
heterogeneous graphs.
Consider for instance a network flow recorded at the border of a protected
network, indicating that a connection has occurred between internal host $u$
and external host $v$.
A collection of such events can be abstracted into a bipartite graph whose
top (resp. bottom) nodes are the internal (resp. external) hosts.
In addition, each network flow is further characterized by additional
categorical variables, such as the transport layer protocol (e.g. TCP or UDP)
or the source and destination ports.
Therefore, each internal--external host pair can be linked by various kinds
of edges.
This suggests that a multiplex graph~\cite{kivela2014multilayer} might be an
appropriate representation.
However, when considering a real-world network, such a graph could easily have
tens of thousands of nodes, making it hard to interpret without any further
processing.

To alleviate this issue, we propose to use a model-based multiplex graph
biclustering algorithm.
This algorithm extends previous work on latent block
modelling~\cite{govaert2003clustering,govaert2010latent} by factoring in the
existence of multiple types of edges.
It allows us to extract relevant clusters of entities along with their
interactions across various network layers, thereby creating a
simplified view of the information contained in the logs.
While graph biclustering has already been used for event log
analysis~\cite{metelli2019bayesian}, to the best of our knowledge, the use of
multilayer models to account for the heterogeneity of events is novel.
In addition, we present detailed case studies on two publicly available
datasets consisting of network flow records and authentication logs,
respectively.

The rest of this paper is structured as follows.
We first introduce the statistical model and inference procedure we use for
multiplex graph biclustering in Section~\ref{sec:model}.
The two case studies are then presented in Section~\ref{sec:flow} and
Section~\ref{sec:auth}, respectively.
Finally, we review some related work in Section~\ref{sec:related} and discuss
potential areas of improvement in Section~\ref{sec:conclusion}.

\section{The Multilayer Latent Block Model}
\label{sec:model}

Before diving into the specifics of authentication logs and network flows,
we first introduce the statistical tools we use to analyze them.
Section~\ref{sec:model:description} presents our generative model for multiplex
bipartite graphs, and Section~\ref{sec:model:inference} describes the
algorithms we use for model inference and selection.

\subsection{Model Description}
\label{sec:model:description}

\paragraph{Key notations.}
Let $\mathcal{U}=\{u_1,\ldots,u_I\},\mathcal{V}=\{v_1,\ldots,v_J\}$ be
two node sets of size $I$ and $J$, respectively.
We consider a multiplex bipartite graph
$\mathcal{G}=(\mathcal{U},\mathcal{V},\mathcal{E})$, where
$\mathcal{E}\subset\mathcal{U}\times\mathcal{V}\times[L]$ denotes
the edge set (with $[L]\overset{\text{def}}{=}\{1,\ldots,L\}$).
Specifically, each edge $(u,v,\ell)\in\mathcal{E}$ between a top node $u$ and
a bottom node $v$ is further characterized by an edge type $\ell$, and $L$
denotes the number of possible edge types.
The multiplex graph $\mathcal{G}$ can then be characterized by $L$ biadjacency
matrices $\mathbf{B}^{(1)},\ldots,\mathbf{B}^{(L)}$, where for each
$\ell\in[L]$, the matrix
\(
	\mathbf{B}^{(\ell)}=
		\big(b_{ij}^{(\ell)}\big)\in\mathbb{R}^{I\times{J}}
\) is defined by \[
	\forall (i,j)\in[I]\times[J],\,b_{ij}^{(\ell)}=\begin{cases}
		1 & \text{ if }(u_i,v_j,\ell)\in\mathcal{E} \\
		0 & \text{ otherwise.}
	\end{cases} 
\]
Finally, we aim to partition $\mathcal{U}$ (resp. $\mathcal{V}$) into a fixed
number $H$ (resp. $K$) of clusters, which we denote
$\mathbb{U}=\{\mathcal{U}_1,\ldots,\mathcal{U}_H\}$
(resp. $\mathbb{V}=\{\mathcal{V}_1,\ldots,\mathcal{V}_K\}$).
For each top node $u_i$ (resp. bottom node $v_j$), the unique index $h$ such
that $u_i\in\mathcal{U}_h$ (resp. $k$ such that $v_j\in\mathcal{V}_k$) is
denoted $U_i$ (resp. $V_j$).
In order to find the optimal partition, we adopt a model-based approach
relying on the generative model described in the next paragraph.

\paragraph{Generative model.}
We propose a multilayer extension of the Poisson latent block model
introduced by Govaert and Nadif~\cite{govaert2010latent}.
This model relies on the fundamental assumption that the probability of an
edge between two nodes depends on the cluster assignments of these nodes.
The biadjacency matrices $\mathbf{B}^{(1)},\ldots,\mathbf{B}^{(L)}$ are then
generated by the following hierarchical model:
\begin{enumerate}[(i)]
	\item for each $i\in[I]$, sample
		$U_i\sim\mathrm{Multinomial}(\boldsymbol{\pi})$;
	\item for each $j\in[J]$, sample
		$V_j\sim\mathrm{Multinomial}(\boldsymbol{\rho})$;
	\item for each $(i,j,\ell)\in[I]\times[J]\times[L]$, sample
		\(b_{ij}^{(\ell)}\sim\mathrm{Poisson}\left(
			\mu_i\nu_j\theta_{U_iV_j}^{(\ell)}
		\right)
		\).
\end{enumerate}
In words, the proportion of top (resp. bottom) nodes falling into each cluster
is controlled by a parameter $\boldsymbol{\pi}\in(0,1)^H$ (resp.
$\boldsymbol{\rho}\in(0,1)^K$).
The probability of an edge of type $\ell$ linking a top node $u_i$ and a
bottom node $v_j$ then depends on three parameters: the node-specific factors
$\mu_i$ and $\nu_j$ represent the overall propensity of each node to form
edges across all layers, and the rate $\theta_{U_iV_j}^{(\ell)}$ controls the
number of edges between clusters $\mathcal{U}_{U_i}$ and $\mathcal{V}_{V_j}$
in the $\ell$-th layer.
Note that the Poisson distribution is only used here to facilitate
calculations: in practice, the biadjacency matrices only contain zeros and
ones.

\subsection{Model Inference and Selection}
\label{sec:model:inference}

Given an observed graph $\mathcal{G}=(\mathcal{U},\mathcal{V},\mathcal{E})$, we
now aim to find the optimal partitions $\mathbb{U}$ and $\mathbb{V}$ using the
model introduced above.
To that end, we need to infer the parameters $\boldsymbol{\pi}$,
$\boldsymbol{\rho}$, $\boldsymbol{\mu}=\big(\mu_i)\in\mathbb{R}_+^{I}$,
$\boldsymbol{\nu}=\big(\nu_j)\in\mathbb{R}_+^{J}$, and
$\mathbf{\Theta}=\big(\mathbf{\Theta}^{(1)},\ldots,\mathbf{\Theta}^{(L)}\big)$
(where
\(\mathbf{\Theta}^{(\ell)}=\big(
	\theta^{(\ell)}_{hk}
	\big)\in\mathbb{R}_+^{H\times{K}}
\) for each $\ell\in[L]$).
The number of top clusters $H$ and bottom clusters $K$ must also be set in
a principled manner.
These two problems are addressed in the next paragraphs.

\paragraph{Model inference.}
The parameters of the model and the cluster assignments are obtained by
maximizing the complete data log-likelihood
\begin{equation*}
\begin{split}
	L_{\mathrm{C}}(\mathbb{T},\mathbb{U},\mathbb{V}) =&
		\sum_{i=1}^{I}\log\pi_{U_i} + \sum_{j=1}^{J}\log\rho_{V_j} \\
		&+ \sum_{i=1}^{I}\sum_{j=1}^{J}\sum_{\ell=1}^L\left\{
			b_{ij}^{(\ell)}\log\left(
				\mu_i\nu_j\theta_{U_iV_j}^{(\ell)}
			\right)
			- \mu_i\nu_j\theta_{U_iV_j}^{(\ell)}
		\right\},
\end{split}
\end{equation*}
where \(\mathbb{T}=\{\boldsymbol{\pi},\boldsymbol{\rho},\boldsymbol{\mu},
\boldsymbol{\nu},\mathbf{\Theta}\}\)
denotes the complete set of parameters.
To that end, we use the block expectation-maximization algorithm described
in~\cite{govaert2010latent}, with minor adjustments to factor in the
multilayer nature of the data.
This algorithm first introduces soft cluster assignment matrices
$\mathbf{U}=\big(\tilde{u}_{ih}\big)\in[0,1]^{I\times{H}}$ and
$\mathbf{V}=\big(\tilde{v}_{jk}\big)\in[0,1]^{J\times{K}}$, with
$\sum_{h=1}^H\tilde{u}_{ih}=\sum_{k=1}^K\tilde{v}_{jk}=1$ for all
$i\in[I]$ and $j\in[J]$.
It then maximizes the fuzzy criterion \[
  G(\mathbb{T},\mathbf{U},\mathbf{V})=
    L_{\mathrm{S}}(\mathbb{T},\mathbf{U},\mathbf{V})
    + H(\mathbf{U}) + H(\mathbf{V}),
\]
where
$H(\mathbf{U})=-\sum_{i=1}^{I}\sum_{h=1}^H\tilde{u}_{ih}\log\tilde{u}_{ih}$
denotes the total entropy of the soft cluster assignments of the top nodes,
$H(\mathbf{V})$ denotes the total entropy for the bottom nodes, and
\begin{equation*}
\begin{split}
  L_{\mathrm{S}}(\mathbb{T},\mathbf{U},\mathbf{V})=&
    \sum_{i=1}^{I}\sum_{h=1}^H\tilde{u}_{ih}\log\pi_h
    + \sum_{j=1}^{J}\sum_{k=1}^K\tilde{v}_{jk}\log\rho_k \\
    &+ \sum_{h=1}^H\sum_{k=1}^K\sum_{i=1}^{I}\sum_{j=1}^{J}\sum_{\ell=1}^L
      \tilde{u}_{ih}\tilde{v}_{jk}\left[
      b_{ij}^{(\ell)}\log\left(
        \theta_{hk}^{(\ell)}
      \right)
      - \mu_i\nu_j\theta_{hk}^{(\ell)}
    \right]
\end{split}
\end{equation*}
is the fuzzy likelihood function.
The criterion $G(\mathbb{T},\mathbf{U},\mathbf{V})$ is maximized by
alternatively performing two steps: optimizing $\mathbf{U}$ and $\mathbf{V}$
with $\mathbf{\Theta}$ fixed (E-step), and optimizing $\mathbf{\Theta}$ with
$\mathbf{U}$ and $\mathbf{V}$ fixed (M-step).
These two steps are iterated until $G$ stabilizes.
The node partitions $\mathbb{U}$ and $\mathbb{V}$ can then be obtained by
assigning each node to the most probable cluster according to $\mathbf{U}$
and $\mathbf{V}$.
Algorithm~\ref{alg:bem} describes the detailed inference procedure.
Note that since the result depends on the random initialization of the
parameters, we run the whole procedure 50 times with different initializations
and return the model with the highest likelihood.

\begin{algorithm}[t]
  \KwData{
    Biadjacency matrices
    $\mathbf{B}^{(1)},\ldots,\mathbf{B}^{(L)}$;
    number of top clusters $H$;
    number of bottom clusters $K$;
    stopping criterion $\epsilon$.
  }
  \KwResult{
    Set of estimated parameters $\mathbb{T}$;
    node partitions $\mathbb{U}$ and $\mathbb{V}$.
  }
  $M\leftarrow\sum_{\ell=1}^L\sum_{i=1}^I\sum_{j=1}^Jb_{ij}^{(\ell)}$\;
  \ForEach{$i\in[I],j\in[J]$}{
    \(
      \mu_i\leftarrow\frac{1}{\sqrt{M}}
        \sum_{\ell=1}^L\sum_{j=1}^J b_{ij}^{(\ell)}
    \);
    \quad
    \(
      \nu_j\leftarrow\frac{1}{\sqrt{M}}
        \sum_{\ell=1}^L\sum_{i=1}^I b_{ij}^{(\ell)}
    \)\;
  }
  Randomly initialize \(
    \mathbf{\Theta},\boldsymbol\pi,\boldsymbol\rho,\mathbf{U},\mathbf{V}
  \);
  \quad
  $\Delta\leftarrow\infty$\;
  \While{$\Delta>\epsilon$}{
    $G_{\mathrm{old}}\leftarrow{G}(\mathbb{T},\mathbf{U},\mathbf{V})$\;
    \ForEach{$i\in[I],h\in[H]$}{
      \(
        s_{ih} \leftarrow \log\pi_h
        + \sum_{k=1}^K\sum_{\ell=1}^L\sum_{j=1}^J
          \tilde{v}_{jk}\left(
            b_{ij}^{(\ell)}\log\theta_{hk}^{(\ell)}
            - \mu_i\nu_j\theta_{hk}^{(\ell)}
          \right)
      \)\;
      \(
        \tilde{u}_{ih} \leftarrow \frac{
          \exp\left(s_{ih}\right)
        }{
          \sum_{h'=1}^H \exp\left(s_{ih'}\right)
        }
      \)\;
    }
    \ForEach{$j\in[J],k\in[K]$}{
      \(
        t_{jk} \leftarrow \log\rho_k
        + \sum_{h=1}^H\sum_{\ell=1}^L\sum_{i=1}^I
          \tilde{u}_{ih}\left(
            b_{ij}^{(\ell)}\log\theta_{hk}^{(\ell)}
            - \mu_i\nu_j\theta_{hk}^{(\ell)}
          \right)
      \)\;
      \(
        \tilde{v}_{jk} \leftarrow \frac{
          \exp\left(t_{jk}\right)
        }{
          \sum_{k'=1}^K \exp\left(t_{jk'}\right)
        }
      \)\;
    }
    \ForEach{$h\in[H],k\in[K]$}{
      $\pi_h\leftarrow\frac{1}{I}\sum_{i=1}^I\tilde{u}_{ih}$;
      \quad
      $\rho_k\leftarrow\frac{1}{J}\sum_{j=1}^J\tilde{v}_{jk}$\;
      \ForEach{$\ell\in[L]$}{
        \(
          \theta_{hk}^{(\ell)} \leftarrow
            \frac{
              \sum_{i=1}^I\sum_{j=1}^J
                \tilde{u}_{ih}\tilde{v}_{jk}b_{ij}^{(\ell)}
            }{
              \left(
                \sum_{i=1}^I \tilde{u}_{ih}\mu_i
              \right)
              \left(
                \sum_{j=1}^J \tilde{v}_{jk}\nu_j
              \right)
            }
        \)\;
      }
    }
    $G_{\mathrm{new}}\leftarrow{G}(\mathbb{T},\mathbf{U},\mathbf{V})$;
    \quad
    \(
      \Delta\leftarrow\left|
        1-G_{\mathrm{new}}/G_{\mathrm{old}}
      \right|
    \)\;
  }
  \KwRet{
    $\mathbb{T},\,\textsc{Round}(\mathbf{U}),\,\textsc{Round}(\mathbf{V})$
  }
  \caption{Block expectation-maximization algorithm for maximum likelihood
    inference of the node clusters and model parameters.
  }
  \label{alg:bem}
\end{algorithm}

\paragraph{Model selection.}
In order to run the aforementioned inference procedure, we must first set the
number of top clusters $H$ and bottom clusters $K$.
In the absence of any prior knowledge, we use the integrated completed
likelihood (ICL~\cite{biernacki2000assessing}) to pick the best values out of
a predefined set of candidates.
More specifically, for each $(H,K)\in\{2,\ldots,16\}^2$, we run the inference
procedure to obtain the optimal parameter set $\mathbb{T}$ and node partitions
$\mathbb{U},\mathbb{V}$.
We then compute \[
  \mathrm{ICL}(\mathbb{U},\mathbb{V};H,K) =
    L_{\mathrm{C}}(\mathbb{T},\mathbb{U},\mathbb{V})
    - \frac{H-1}{2}\log I - \frac{K-1}{2}\log J
    - \frac{LHK}{2}\log(LIJ)
\]
for each candidate model, and the highest-scoring model is selected.
The ICL penalizes the complete data log-likelihood by subtracting terms
proportional to the number of free parameters, thus enabling a trade-off
between goodness-of-fit and model complexity.
Note that it relies on several approximations, which makes it an arguably
imperfect evaluation criterion.
This has motivated many subsequent contributions on model selection for
stochastic and latent block models (see
e.g.~\cite{come2015model,lomet2012model,wyse2017inferring}).
However, for the sake of simplicity, we leave the use of more sophisticated
criteria for future work.

\section{First Case Study --- Network Flows}
\label{sec:flow}

Having described our modelling tools, we now move on to our first case study.
Section~\ref{sec:flow:data} describes the dataset we consider and the
preprocessing steps we apply to turn it into a multiplex network.
We then discuss the results obtained by applying our methodology in
Section~\ref{sec:flow:results}.

\subsection{Data Description}
\label{sec:flow:data}

The dataset we use was originally created for the Mini-Challenge 3 of the VAST
2013 competition~\cite{whiting2013vast}.
It represents two weeks of simulated network traffic between approximately
1\,400 hosts, most of which belong to an enterprise network.
We refer to these hosts as internal ones, while hosts that are not part of the
enterprise network are called external hosts.
The dataset contains benign traffic as well as various kinds of attacks.
Some of these attacks are rather noisy --- e.g. distributed denials of service
(DDoS), port scans --- while others are more subtle (e.g. data exfiltration).
Overall, 53 out of 200 external hosts are involved in malicious traffic, which
is a rather high ratio.
This, along with the small number of hosts and synthetic nature of the data,
makes this dataset a somewhat easy test case for our methodology.
Moreover, a thorough description of the hosts and their respective roles is
available, as well as a complete list of attack-related events.
This allows us to evaluate the relevance of our results.

Note that while the dataset encompasses several data sources, we only keep
the network flow records between internal hosts and external hosts.
We turn these records into a bipartite multiplex graph as follows.
First, each internal (resp. external) host is represented by a top (resp.
bottom) node.
For each flow between an internal host \textsf{IH} and an external host
\textsf{EH}, we then build an edge between \textsf{IH} and \textsf{EH}, of type
(\textit{Protocol}, \textit{Destination Port}, \textit{Direction}).
Note that due to the large number of possible destination ports, we only keep
10 distinct values corresponding to well-known protocols\footnote{TCP/20
(FTP-Data), TCP/21 (FTP), TCP/22 (SSH), TCP/23 (Telnet), TCP/25 (SMTP), TCP/53
(DNS), TCP/80 (HTTP), TCP/443 (HTTPS), TCP/465 (SMTPS), and
TCP/587 (SMTP message submission).} and
represent the other values as a single \textsf{Other} token.
As for the direction of the flow, it is either inbound (the internal host is
the destination) or outbound (the internal host is the source).
Note that each observed (\textsf{IH}, \textsf{EH}, \textsf{Type}) triple is
represented by one single edge, regardless of its number of occurrences.
See Table~\ref{tab:datasets} for some descriptive statistics about the dataset
and the obtained graph.

{
\setlength{\tabcolsep}{1em}
\begin{table}[t]
  \caption{Description of the datasets:
    number of top nodes $I$,
    number of bottom nodes $J$,
    number of layers $L$,
    number of distinct edges $M$,
    and total number of events $N$.
  }
  \centering
  \begin{tabular}{lrrrrr}
    \toprule
    \textbf{Dataset} & $\boldsymbol{I}$ & $\boldsymbol{J}$ & $\boldsymbol{L}$
    & $\boldsymbol{M}$ & $\boldsymbol{N}$ \\
    \midrule
    VAST & 1\,220 & 200 & 18 & 26\,597 & 68\,793\,510 \\
    LANL & 74\,049 & 16\,119 & 44 & 869\,547 & 842\,282\,832 \\
    \bottomrule
  \end{tabular}
  \label{tab:datasets}
\end{table}
}

\subsection{Results}
\label{sec:flow:results}

\begin{figure}[t]
  \centering
  \includegraphics[width=\textwidth]{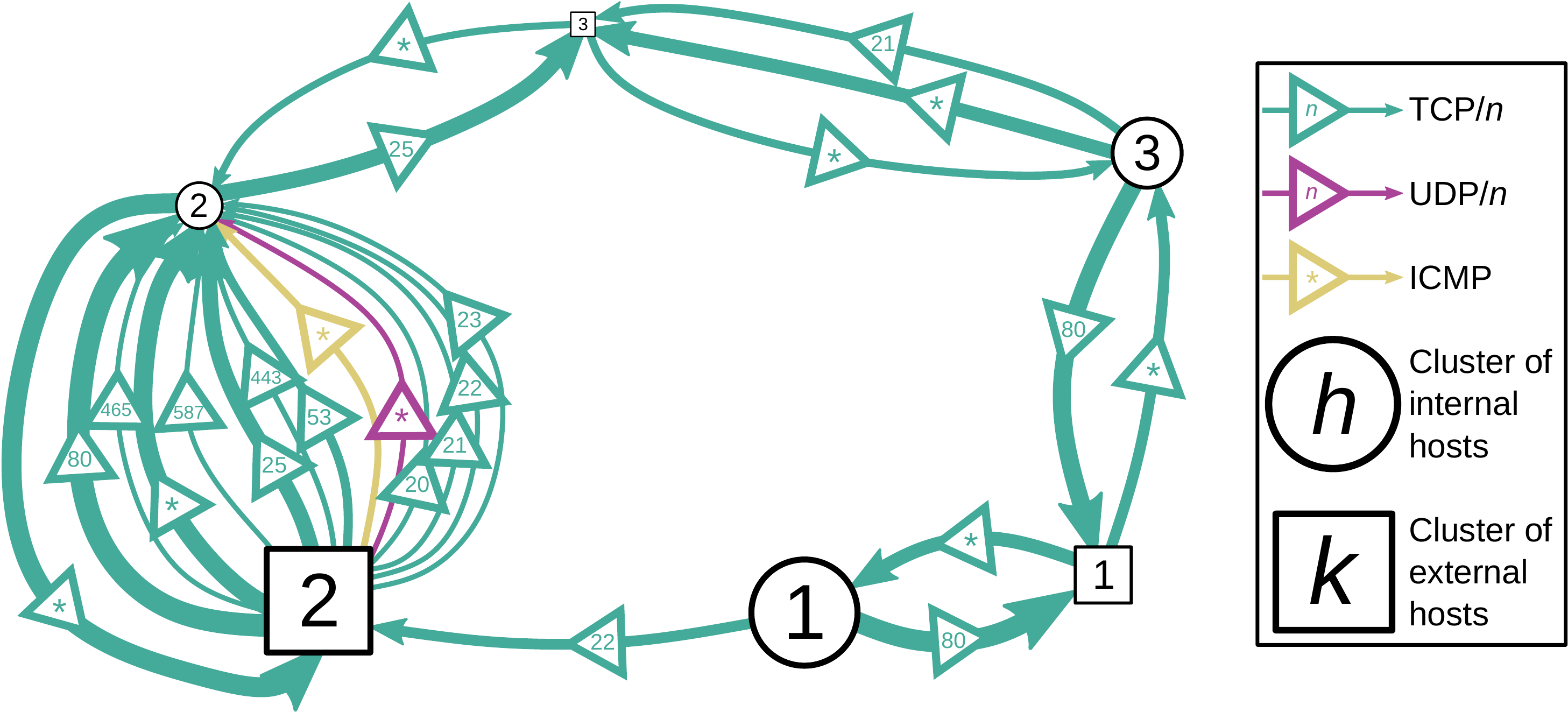}
  \caption{Clusters found in the VAST dataset and interactions
    between them.}
  \label{fig:vast}
\end{figure}

Our methodology yields three clusters of internal hosts and three clusters of
external hosts.
These clusters, as well as the aggregated edges between them, are shown in
Figure~\ref{fig:vast}.
Note that for the sake of readability, only edges representing at least 40
events are displayed.
The size of each cluster is proportional to the logarithm of the number of
hosts it contains.
Similarly, the width of each edge grows logarithmically with the number of
underlying events.

Starting with internal clusters, we observe that cluster 1 initiates HTTP
and SSH connections, and that it has more outbound flows than inbound ones.
Similarly, cluster 3 primarily interacts through outbound flows, including
HTTP and FTP traffic.
In contrast, cluster 2 receives many inbound connections, with HTTP, SMTP
and DNS among the most represented protocols.
It also sends SMTP traffic to external cluster 3.
Overall, we can thus safely assume that clusters 1 and 3 contain workstations
while cluster 2 contains servers, which is indeed confirmed by the ground
truth description.
As for external clusters, cluster 1 mostly receives HTTP connections from
internal workstations, suggesting that it contains Web servers.
This is consistent with the ground truth.
Cluster 3 contains a mail server and an FTP server, which explains the
inbound SMTP traffic coming from internal servers as well as the inbound FTP
connections coming from internal workstations.
Finally, cluster 2 contains the majority of external hosts, which are
primarily observed connecting to the internal servers.
In particular, attackers all fall into cluster 2, along with many benign
hosts.

Even though partitioning the set of external hosts does not isolate malicious
ones, studying the interactions between clusters allows us to uncover several
attacks.
For instance, all TCP traffic from external cluster 2 to internal cluster 2 on
ports 20, 21, 22, 23, 53, 443, 465 and 587 originates from two hosts (10.9.81.5
and 10.10.11.15), which happen to be attackers.
This also holds for all UDP traffic between these two clusters.
These edges stand out in Figure~\ref{fig:vast} because of the relatively small
number of events they represent.
Such scarce connections on many different ports suggest port scanning
activity, which is confirmed by the ground truth.
Similarly, ICMP traffic from external cluster 2 to internal cluster 2 also
results from network scans.
Finally, all SSH traffic from internal cluster 1 to external cluster 2 is
directed towards one single host (10.0.3.77), which calls for further
investigation.
These flows actually represent beaconing activity from compromised internal
hosts to a command and control server.

Overall, this first case study shows that our approach can indeed infer
meaningful clusters, thereby revealing the functional roles of most hosts.
It can also help identify some malicious behaviors by providing a reduced
number of starting points for deeper investigation.
Note, however, that some types of attacks are not easily detectable.
Typically, data exfiltrations and DDoS attacks only differ from normal traffic
through their volume.
Since our graph-based representation does not include this characteristic, it
does not allow us to distinguish exfiltrations from regular FTP traffic, or
DDoS attacks from regular Web traffic.
However, these high-volume attacks can be detected using other methods.

\section{Second Case Study --- Authentication Logs}
\label{sec:auth}

Our second case study focuses on a different data source, namely authentication
logs collected within a real-world enterprise network.
The dataset, which we describe in Section~\ref{sec:auth:data}, is significantly
larger and more complex than the one studied in the previous section.
It is thus more challenging to extract meaningful insights from it.
However, as we show in Section~\ref{sec:auth:results}, our methodology still
allows us to infer some functional roles and uncover malicious behavior.

\subsection{Data Description}
\label{sec:auth:data}

We use the "Comprehensive, Multi-Source Cyber-Security 
Events"~\cite{kent2015cyberdata,kent2015cybersecurity} dataset released by the
Los Alamos National Laboratory (LANL).
This dataset represents 58 days of activity within the LANL's enterprise
network, recorded through several kinds of event logs.
A red team exercise took place during this time span, meaning that penetration
testers tried to breach the network in order to assess its security.
The remote authentications performed by the red team are labelled, providing
and interesting example of an advanced intrusion within an enterprise network.
In particular, this attack is stealthier than those considered in the
previous section, which makes it a more challenging test case for our method.

We focus on Windows authentication logs, more specifically
on successful \textit{LogOn} events.
Each one of these events is described by a source user \textsf{SU}, a
destination user \textsf{DU}, a source host \textsf{SH}, a destination host
\textsf{DH}, a logon type \textsf{LT} and an authentication package
\textsf{AP}.
Note that \textsf{SU} and \textsf{DU} can be identical, and the same goes
for \textsf{SH} and \textsf{DH}.
As for \textsf{LT} and \textsf{AP}, they describe the type of session being
created and the protocol used to authenticate the user, respectively.
Examples of logon types include Interactive (local session with graphical
interface) or Network (used for remote file accesses or remote procedure calls,
for instance), and
the most frequent authentication packages are Kerberos and NTLM.
Note that these two additional fields are especially meaningful as they allow
to distinguish different behaviors (e.g., remote administrative session versus
simple file access).
They can also help characterize users and hosts: for instance, since NTLM is
considered less secure than Kerberos, NTLM authentication can sometimes be
disabled for highly privileged accounts.
Conversely, some legacy applications may not support Kerberos authentication,
thus servers hosting such applications should frequently use NTLM.

We represent users as top nodes and hosts as bottom nodes, and we turn logon
events into typed edges as follows.
If \textsf{SH} and \textsf{DH} are identical, we create one edge between
\textsf{DU} and \textsf{DH} with type
(\textsf{LT}, \textsf{AP}, \textsf{Local}).
Otherwise, we create one edge between \textsf{SU} and \textsf{SH} with
type (\textsf{LT}, \textsf{AP}, \textsf{From}), and one edge between
\textsf{DU} and \textsf{DH} with type (\textsf{LT}, \textsf{AP},
\textsf{To}).
While this construction breaks the link between the source and destination
of an authentication, it still highlights the difference between hosts
receving many remote authentications (typically servers) and hosts from
which these authentications originate (typically workstations).
See Table~\ref{tab:datasets} for some descriptive statistics about the dataset.

\subsection{Results}
\label{sec:auth:results}

\begin{figure}[t]
  \centering
  \includegraphics[width=\textwidth]{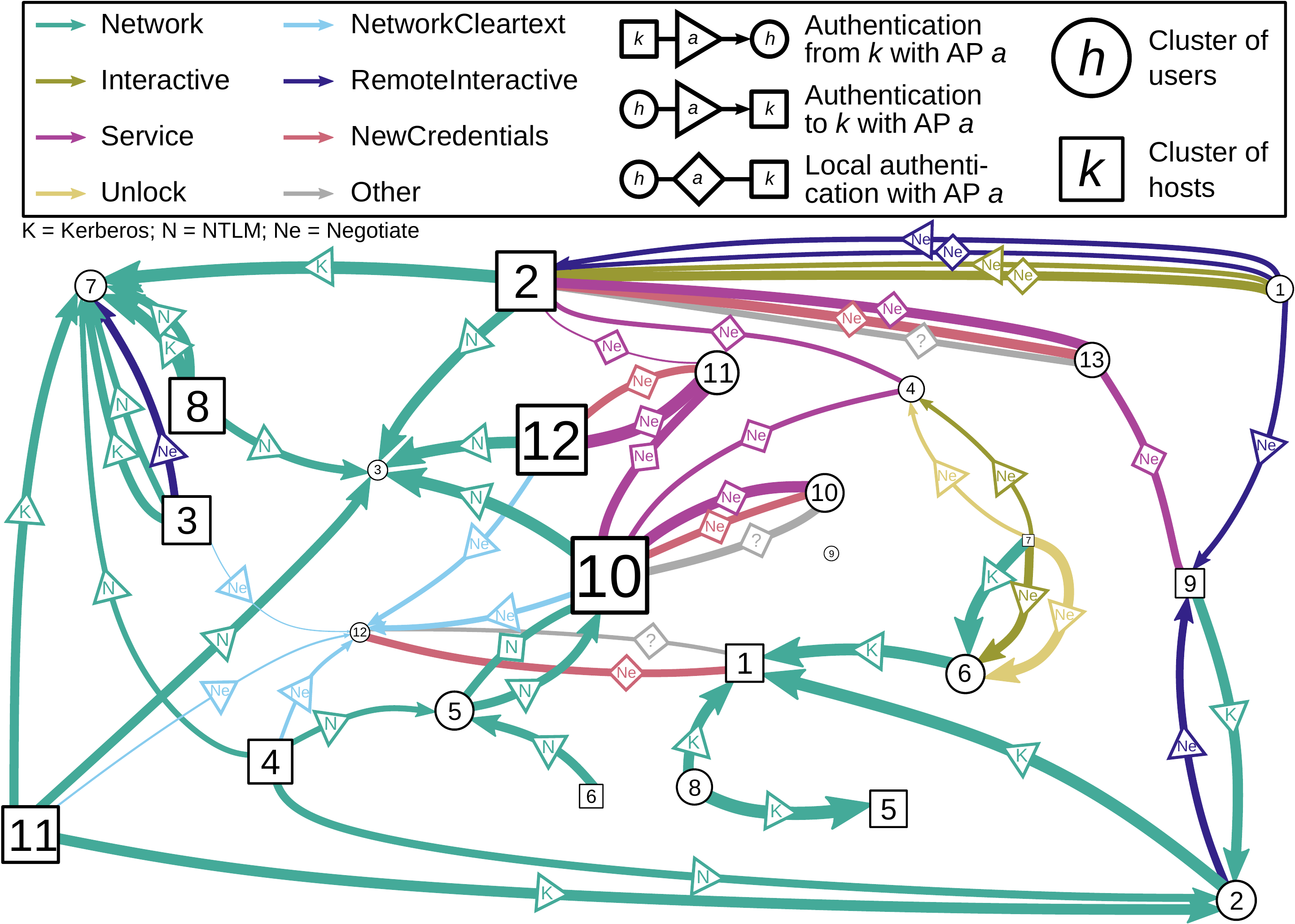}
  \caption{Clusters found in the LANL dataset and interactions
    between them.}
  \label{fig:lanl}
\end{figure}

The optimal model has thirteen user clusters and twelve host clusters.
We display them in Figure~\ref{fig:lanl}, along with the main interactions
between them.
Specifically, we display all edges $(h,k,\ell)$ such that
$\theta_{hk}^{(\ell)}\geq{0}.7$.
Note that we select edges using the set of rate matrices $\mathbf{\Theta}$
rather than the number of underlying events because the behaviors we seek to
detect generate relatively few events.
Thus cluster--cluster edges with high rate, which represent many user--host
edges occurring at least once, are more relevant than those representing many
individual events.

The global picture is expectedly more complex than in the previous section.
In addition, little information is available on the true functional roles of
users and hosts.
However, we can still infer the meaning of some clusters and confirm their
relevance.
For instance, user clusters 10, 11 and 13 mainly interact through local
authentications with logon type Service, indicating that they contain service
accounts.
User cluster 1 opens remote interactive sessions on many hosts, which suggests
the presence of administrator accounts.
As for user clusters 3 and 5, they mostly consist of anonymous user
credentials, which are typically used to access shared resources which do not
require any access control (such as internal Web pages).
This is consistent with the NTLM-authenticated network logons associated with
these clusters.
Interestingly, a significant proportion of the non-anonymous user names from
these two clusters were used for red team activity.
This suggests that looking for inconsistencies between the expected functional
role of an entity and the cluster it falls into can help uncover malicious
behaviors.

Inferring the functional roles of host clusters is more difficult, although it
seems reasonable to assume that cluster 1 contains domain controllers and
applicative servers.
Indeed, this cluster is the one receiving the most remote authentications.
Conversely, clusters 2, 3, 8, 9, and 11 are sources of remote authentications,
suggesting that they mostly contain workstations.
Finally, cluster 4 stands out as the source of remote NTLM authentications
involving three user clusters.
Further investigation reveals that most of these authentications were performed
by the red team and originate from host C17693, which also happens to be the
main source of red team activity.
Once again, the simplified situational picture we generate thus preserves
important clues that can lead to the detection of malicious behavior.

\section{Related Work}
\label{sec:related}

\paragraph{Biclustering and latent block models.}
Biclustering, i.e., the idea of simultaneously partitioning the rows and
columns of a matrix so as to form blocks of coefficients with similar values,
can be traced back to the early 1970s~\cite{hartigan1972direct}.
Spectral methods~\cite{dhillon2001co,kluger2003spectral} are among the most
popular approaches to this task.
However, model-based biclustering has gained traction in the last two decades,
following the introduction of the latent block
model~\cite{govaert2003clustering}.
Various inference procedures have been proposed, most of which rely on the
expectation-maximization (EM) algorithm.
While we apply the simple variational approximation proposed
in~\cite{govaert2003clustering}, other approaches include the addition of
intermediary classification steps~\cite{govaert2008block} or the use of
Bayesian inference~\cite{keribin2015estimation}.

\paragraph{Multilayer stochastic block models.}
Even though we are not aware of any previous work on multilayer generalizations
of the latent block model, a straightforward connection can be made with
multiplex stochastic block models (SBMs).
More specifically, our model is analogous to the degree-corrected SBM with
independent layers described in~\cite{peixoto2015inferring}.
Note, however, that many other multilayer generalizations of the SBM can be
found in the literature.
In particular, the existence of diversely complex statistical dependencies
between layers has been addressed in various
ways~\cite{barbillon2017stochastic,de2017community,paul2016consistent,%
stanley2016clustering}.

\paragraph{Data exploration and visualization for cybersecurity.}
Finally, our work contributes to a vast research effort aiming to ease network
security monitoring by providing insightful visualizations.
In particular, displaying network flows between internal and external hosts is
a recurring challenge.
VISUAL~\cite{ball2004home}, VisFlowConnect~\cite{yin2004visflowconnect},
NFlowVis~\cite{fischer2008large}, and FloVis~\cite{taylor2009flovis} are some
of the existing tools designed to perform this task.
However, none of them focuses on statistical modelling to build a more
condensed view.
Note that FloVis and NFlowVis use the hierarchical nature of IP addresses to
aggregate hosts by subnetwork, and NFlowVis also uses $k$-medoids clustering
to find external hosts with similar communication patterns.
Even so, the use of behavior-based biclustering makes our method more effective
at reducing the number of elements to display while preserving essential
information.
As for authentication logs, APTHunter~\cite{siadati2016detecting} also focuses
on interactivity rather than automated data reduction, only letting the user
manually apply filters to make the authentication graph legible.
Finally, the work of Glatz et al.~\cite{glatz2014visualizing} adopts an
approach somewhat similar to ours: they first extract frequent itemsets from
network flow records, then display them as a bipartite itemset--item graph.
While this does provide a simple summary of large volumes of logs, the
extracted itemsets still cover a small minority of the total number of events,
leading to significant information loss.

\section{Conclusion and Perspectives}
\label{sec:conclusion}

We propose a graph-oriented approach to event log exploration for network
security monitoring.
Through the use of model-based multiplex graph biclustering, we aim to extract
meaningful clusters of entities, such as groups of users or hosts sharing
functional roles or behavioral patterns.
Our case studies demonstrate that such meaningful clusters can indeed be
uncovered.
In addition, displaying interactions between these clusters can facilitate
malicious behavior detection.

Aside from investigating better model selection criteria (see
Section~\ref{sec:model:inference}), extending our model to factor in the
temporal dimension could be an interesting lead for future work.
Previous contributions on temporal latent block models~\cite{corneli2015exact}
provide foundations for such an extension.
Finally, looking for meaningful groups of edge types in addition to entity
clusters, similarly to the strata multilayer SBM~\cite{stanley2016clustering},
is another promising direction.
Indeed, it could lead to an even simpler situational picture, especially for
large and complex datasets such as the one studied in Section~\ref{sec:auth}.

%
%

\bibliographystyle{splncs03}
\bibliography{references}

%


\end{document}